\documentclass[11pt]{article}
\usepackage{moriond,epsfig}

\def\be{\begin{equation}}
\def\ee{\end{equation}}
\def\bea{\begin{eqnarray}}
\def\eea{\end{eqnarray}}

\begin{document}
\vspace*{4cm}
\title{NEUTRINO SELF INTERACTIONS IN SUPERNOVAE}

\author{Gianluigi Fogli$^1$, 
		Eligio Lisi$^1$, 
		Antonio Marrone$^{1},$\footnote[1]{Speaker}
 		Alessandro Mirizzi$^{1,2}$ }
\address{$^1$~Dipartimento di Fisica and Sezione INFN di Bari,
         Via Amendola 173, 70126 Bari, Italy\\ $^2$~Max-Planck-Institut f\"ur Physik
(Werner-Heisenberg-Institut), F\"ohringer Ring 6, 80805 M\"unchen,
Germany}

\maketitle\abstracts{
Oscillations of neutrino emerging from a supernova core are studied.
In this extremely high density region neutrino self interactions induce
collective flavor transitions. When collective transitions are decoupled 
from matter oscillations, as for our chosen matter profile,
an analytical interpretation of the collective effects is possible,
by means of a mechanical analogy with a spherical pendulum.
For inverted neutrino hierarchy the neutrino propagation can be divided in three 
regimes: synchronization, bipolar oscillations, and spectral split. 
Our simulation shows that averaging over neutrino trajectories
does not alter the nature of these three regimes.}

\section{Introduction}

Supernova neutrino oscillations are a very important tool to study astrophysical
processes and to better understand neutrino properties~\cite{Raffelt:2007nv}. 
When neutrinos leave the surface of the neutrinosphere, they undergo vacuum
and matter oscillations. Beside this, in the first few hundred kilometers
neutrino-neutrino interactions induce collective flavor transitions, whose effect
can be very important, depending on the neutrino mass hierarchy.
Self-interaction effects are expected to be non negligible when
$\mu(r) \sim \omega$, where
$\mu(r)$ is the neutrino potential associated to the neutrino background 
($\mu =  \sqrt{2} G_F (N_\nu(r) + \overline{N}_\nu(r))$, 
analogously to the MSW potential $\lambda = \sqrt{2} G_F N_{e^{-}}(r)$ )
and $\omega$ is the vacuum oscillation frequency.
We neglect the solar mass square difference
$\delta m^2=m^2_2-m^2_1\ll \Delta m^2=|m^2_3-m^2_{1,2}|$, and consider a two-neutrino mixing 
scenario where the oscillations are governed by the mixing angle $\theta_{13}$.
Since in the supernova context $\nu_\mu$ and $\nu_\tau$ cannot be distinguished
we generically speak of $\nu_e \leftrightarrow \nu_{x}$ oscillations.
In our work we assume $\Delta m^2 = 10^{-3}$~eV$^2$ and $\sin^2\theta_{13} = 10^{-4}$.
Figure 1 shows the radial profiles of the matter potential $\lambda(r)$ and of
the neutrino potential $\mu(r)$, and the approximate ranges where 
collective flavor transitions of different type occur: synchronization,
bipolar oscillation and spectral split. 
The nonlinearity of the self interactions induce neutrino oscillations very different 
from the ordinary MSW effect. When undergoing collective flavor transition
neutrinos and antineutrinos of any energy behave similarly, as we will see in the following.
This kind of transitions occurs for small $r$, well before  the ordinary
MSW resonance, allowing for a clear interpretation of the numerical simulations.
For matter profiles different from our own,
the MSW resonance condition can occur in the same region 
of the collective transitions: shallow electron density profiles~\cite{Duan:2006an} 
can trigger MSW effects around $O(100)$~km. In that case it is much
more difficult to disentangle collective from MSW effects in the results of
the simulations.

\begin{figure}
\hspace{0.85cm}
\psfig{figure=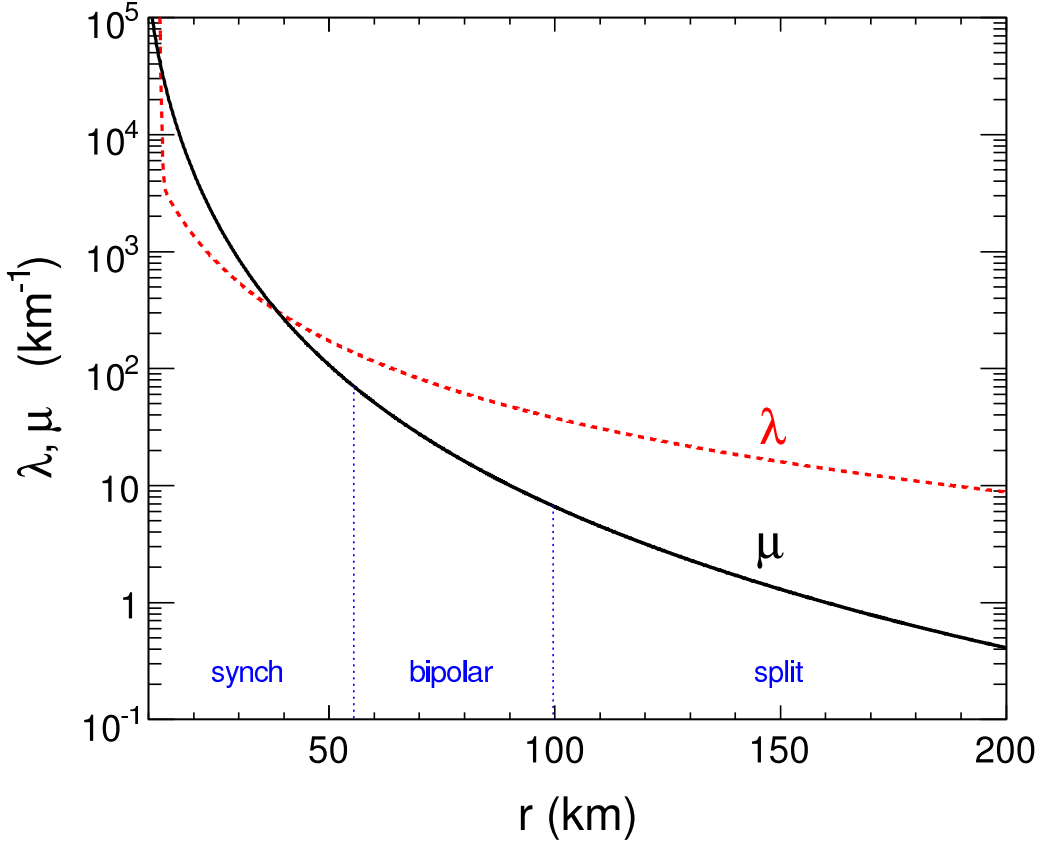,height=4.4in}
\caption{Radial profiles of the neutrino self-interaction parameter
$\mu(r)=\sqrt{2}\,G_F\,(N+\overline N)$ and of the matter-interaction
parameter $\lambda(r)=\sqrt{2}\,G_F\,N_{e^-}$ adopted in this work,
in the range $r\in[10,\,200]$~km. 
\label{fig2}}
\end{figure}

\section{Reference model and pendulum analogy}

In our work, we use normalized thermal spectra with $\langle E_e \rangle = 10$~MeV, 
$\langle \overline E_e \rangle = 15$~MeV, 
and $\langle E_x \rangle = \langle \overline E_x \rangle = 24$~MeV
for $\nu_e$, $\overline\nu_e$, $\nu_x$ and $\overline \nu_x$,
respectively. 
The geometry of the model, the so called
 ``bulb model''~\cite{Duan:2006an}, has a spherical symmetry, since we assume
 that neutrinos are half-isotropically emitted from the neutrinosphere. 
Along any radial trajectory there is, therefore, a cylindrical symmetry.
By virtue of that, we need only two
independent variables to describe the neutrino propagation and interaction:
the distance form the supernova center $r$, and the angle $\vartheta$ between two interacting neutrinos.
If the dependence on $\vartheta$ is
integrated out, we speak of ``single-angle''  approximation, while the general situation
of variable $\vartheta$ is dubbed ``multi-angle'' case. The numerical simulation in the multi-angle case
is extremely challenging, since it requires the solution of a large system (size of order $10^{5}$) of coupled non-linear equations.
\begin{figure}[t]
\begin{minipage}{18pc}
\psfig{figure=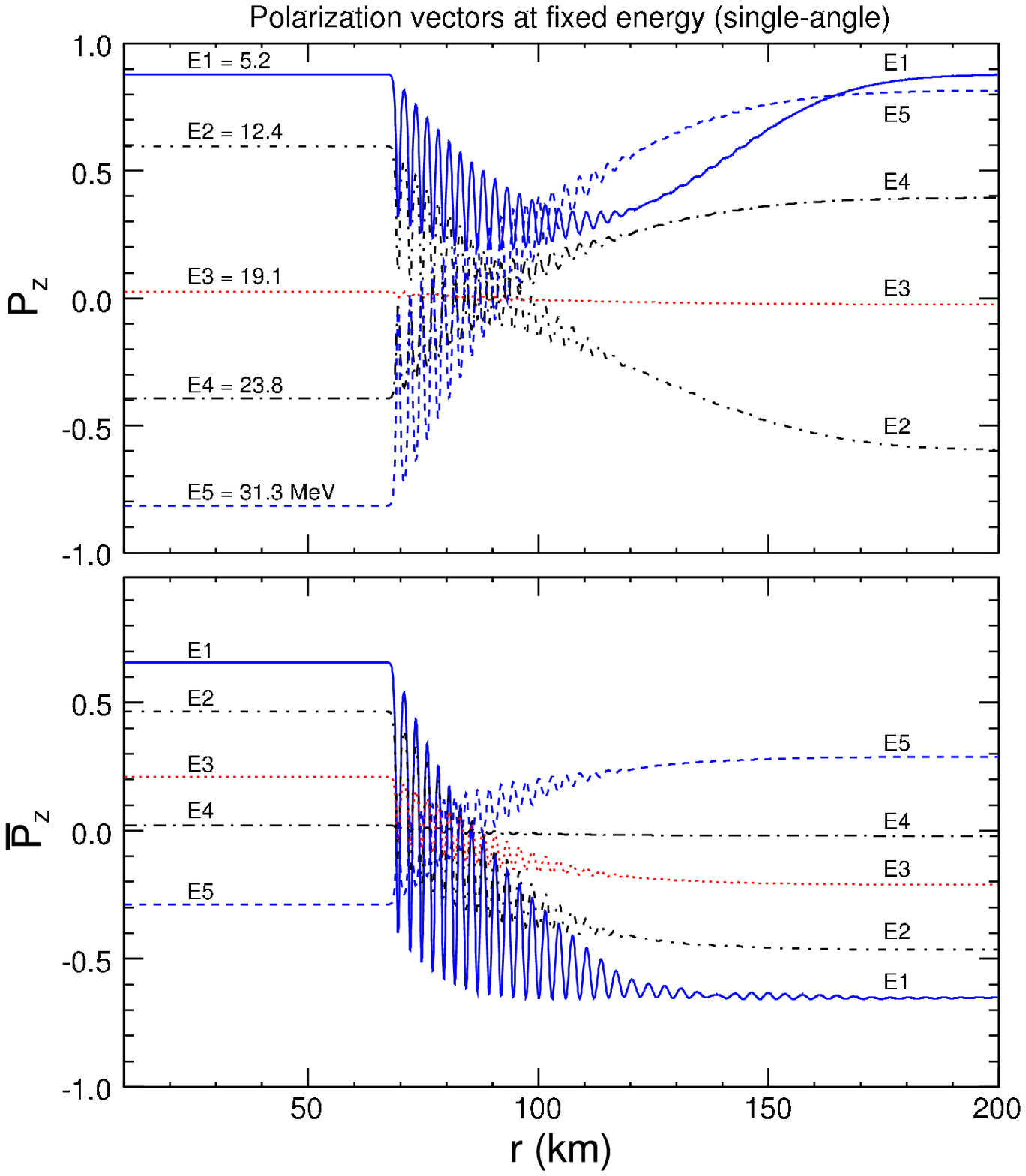,height=3.4in}
\caption{Single-angle simulation in inverted hierarchy: $P_z$ (neutrinos)  and $\overline P_z$ (antineutrinos)
as a function of radius, for five energy values.
\label{fig4}}\end{minipage}
\hspace{1pc}%
\begin{minipage}{18pc}
\psfig{figure=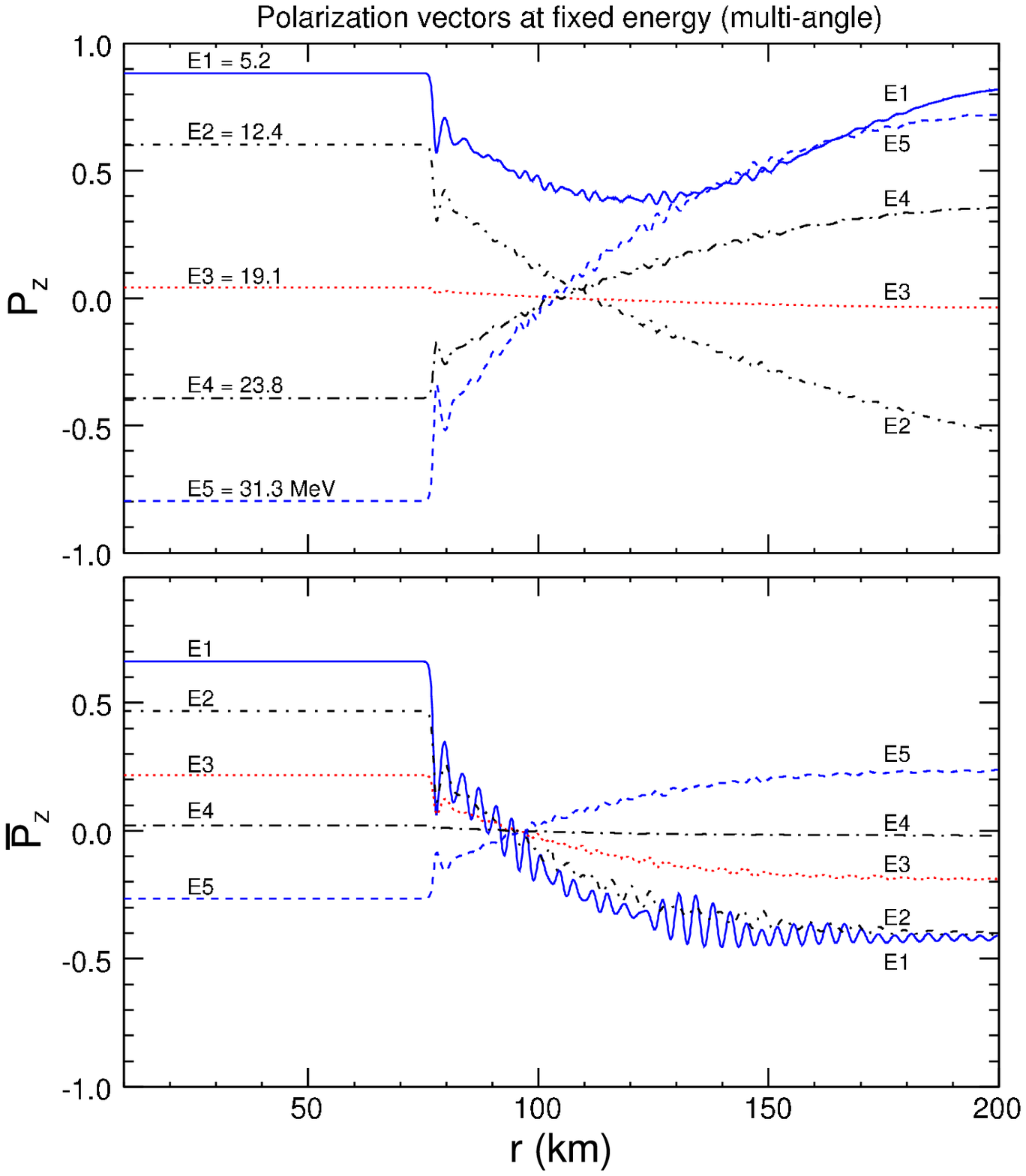,height=3.4in}
\caption{Multi-angle simulation in inverted hierarchy: $P_z$ (neutrinos)  and $\overline P_z$ (antineutrinos)
as a function of radius, for five energy values.
\label{fig7}}
\end{minipage} 
\end{figure}
The propagation of neutrinos of given energy $E$ is studied through the Liouville
equation for the density matrix. 
By expanding the density matrix on the Pauli matrices and on the identity, the equations of motion
can be expressed in terms of two polarization vectors, ${\bf P}(E)$ and $\bar{\bf P}(E)$, for neutrinos
and antineutrinos, respectively. By introducing a vector ${\bf B}$ that depends on the mixing angle 
$\theta_{13}$, and a vector ${\bf D }= {\bf J - \overline{J}}$ that is the difference between the integral
over the energy of ${\bf P}$ and ${\bf \overline P}$, the equations of motion can be written as
\begin{eqnarray}
\dot \mathbf{P}&=&\left(+\omega {\bf B}+\lambda {\bf z}+\mu {\bf D}\right)\times {\bf P}\ ,\label{Bloch1}\\
\dot \mathbf{\overline P}&=&\left(-\omega {\bf B}+\lambda {\bf z}+\mu {\bf D}\right)\times {\bf\overline P}\ .\label{eqm}
\end{eqnarray}
In the general case, the polarization vectors depend also on the
neutrino emission angle $\theta_{0}$ (the neutrino incidence angle $\vartheta$ can be expressed in terms of $r$ 
and of the emission angle at the neutrinosphere $\theta_{0}$).
The electron neutrinos survival probability
$P_{ee}$ is a function of the polarization vector,
$P_{ee}=1/2(1+P^z_f/P_z^i)$, where the $i$ and $f$ refer to the initial and final
state respectively (analogously for antineutrinos).
The equations of motion for  ${\bf P}(E)$ and $\bar{\bf P}(E)$ can be reduced (under
reasonable approximations~\cite{ourCollective}) to the equations of motion of a gyroscopic pendulum,
a spherical pendulum of unit length in a constant gravity field, characterized by a point-like massive bob spinning around the pendulum axis with constant angular momentum.
The pendulum inertia is inversely proportional to $\mu(r)$, while its
total angular momentum depends on the difference of the integrated polarization
vectors $\bf J$ and $\bf \bar J$~\cite{ourCollective}. 
The motion of a spherical pendulum is, in general, a combination of a precession
and a nutation~\cite{Hannestad:2006nj,Duan:2007mv}. 
In the case of normal hierarchy of the neutrino mass spectrum 
the pendulum starts close to the stable, downward 
position and stays close to it, as $\mu$ slowly decreases and no collective effect is present.
In the inverted hierarchy case, the pendulum starts close to the ``unstable,''
upward position.
At the beginning, for small $r$, when $\mu$ is large ($m$ is small), the 
bob spin dominates  and the pendulum remains precessing in the
upward position conserving angular momentum~\cite{Duan:2007mv}, a
situation named synchronization ~\cite{Pastor:2001iu,Hannestad:2006nj}.
Nevertheless, since $\mu$ decreases with $r$, at a certain point 
any $\theta_{13}\neq 0$ triggers the fall of the pendulum and its subsequent  
nutations, the so called bipolar oscillations.
The increase of the pendulum inertia with $r$ reduces the amplitude of the nutations, and
bipolar oscillations are expected to vanish when self-interaction 
and vacuum effects are of the same size. At this point,
at the end of the bipolar regime, 
self-interaction effects do not completely vanish and the spectral split 
builds up:  a ``stepwise swap'' between the $\nu_e$ and $\nu_x$
energy spectra. The neutrino swapping can be explained by the conservation of the pendulum energy
and of the lepton number~\cite{RaffSmirn}.
The lepton number conservation is related to  the constancy of $D_z=J_z-\overline J_z$,
that is a direct consequence of the equation of motion.
For a detailed description of the pendulum analogy and of our reference model
the reader is referred to our previous work~\cite{ourCollective} and references therein.
\begin{figure}[t]
\begin{minipage}{18pc}
\psfig{figure=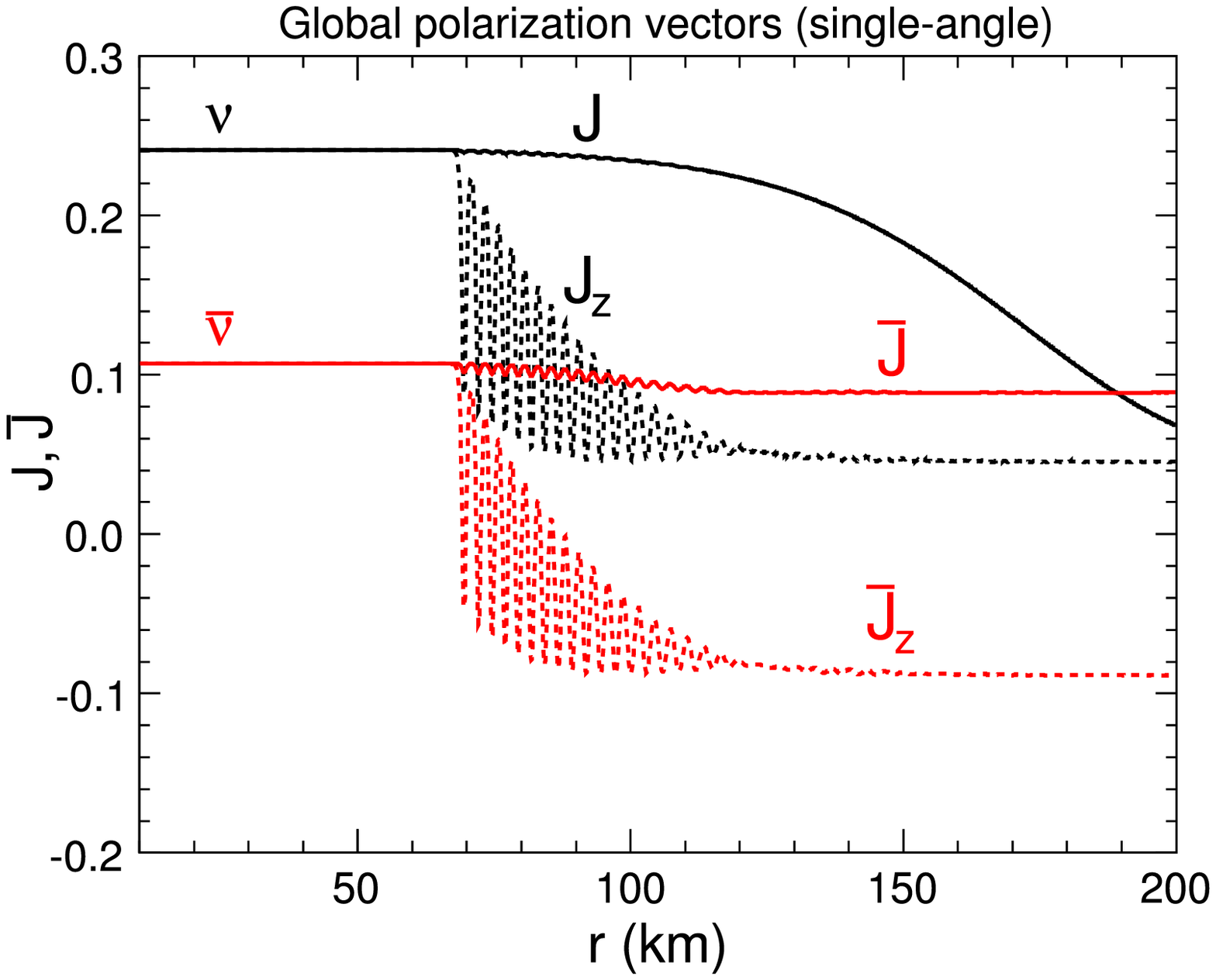,height=2.4in}
\caption{Single-angle simulation in inverted hierarchy: modulus and $z$-component of
$\bf J$ and $\overline{\bf J}$.
\label{fig3}}\end{minipage}
\hspace{2pc}%
\begin{minipage}{18pc}
\psfig{figure=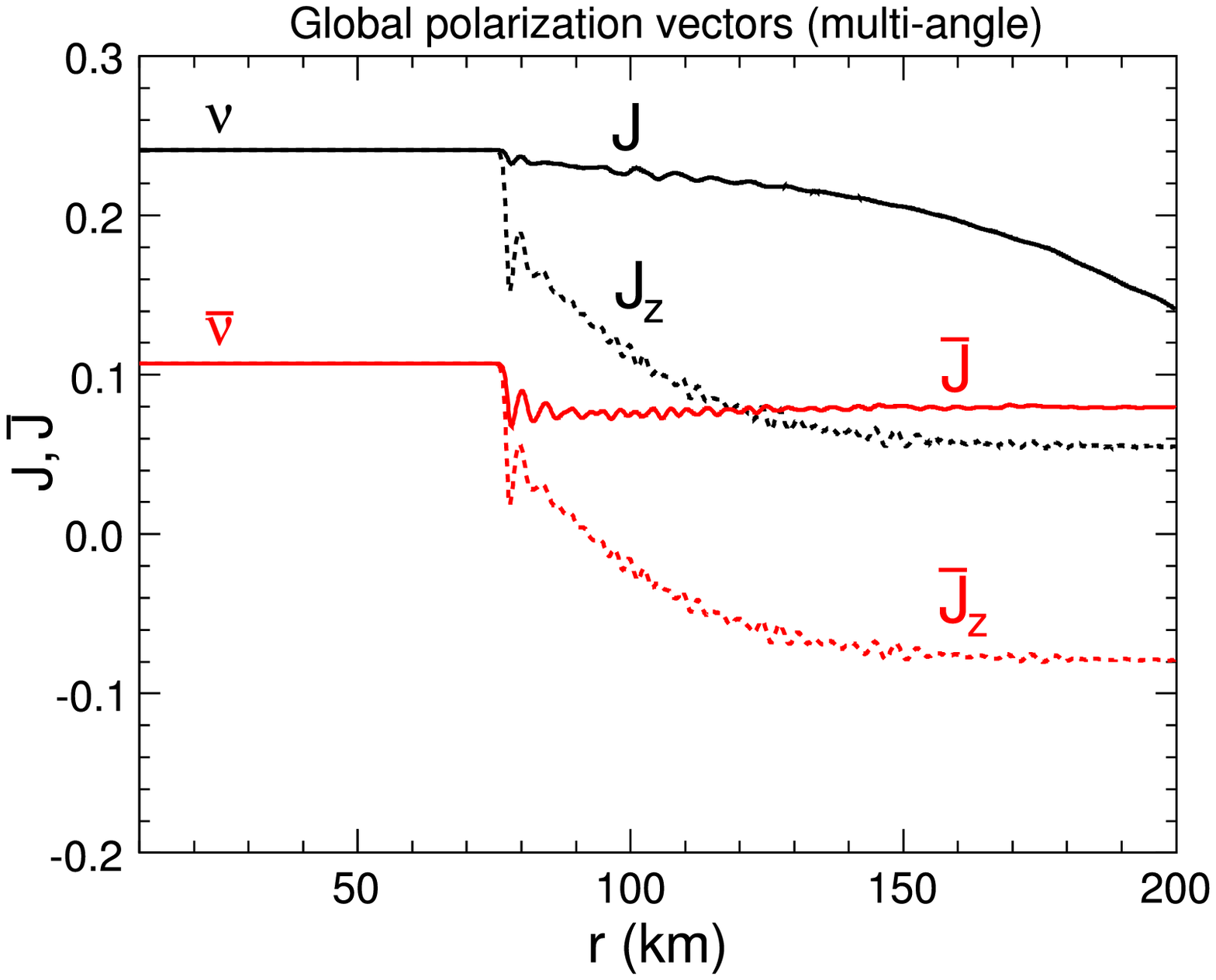,height=2.4in}
\caption{ Multi-angle simulation in inverted hierarchy: modulus and $z$-component of $\bf J$ 
and $\overline{\bf J}$.
\label{fig6}}
\end{minipage} 
\end{figure}

\section{Simulations}

Figures 2 and 3 show the third component of ${\bf P}$ and $\bar{\bf P}$, as a function of the radius,
for different energy values, for the single- and multi-angle simulations, respectively.
Bipolar oscillations starts at the same $r$ and their periods are equal for both $\nu$
and $\overline\nu$ at any energy, confirming the appearance  of a self-induced collective behavior, in the single- and in the multi-angle case. The behavior of each $P_{z}$ and $\overline{P}_{z}$ depends on its energy.
For neutrinos, Figure 2, the spectral split starts around the critical energy
$E_c\simeq 7$ MeV: the curve relative to $E<E_c$ ends up at the same initial value ($P_{ee} =1$),
while the curves for $E>E_c$ show the $P_z$
inversion ($P_{ee} = 0$). 
Neutrinos with an energy of $\sim 19$ MeV do not oscilate much, because this is roughly the energy for which
the initial $\nu_{e}$ and $\nu_{x}$ fluxes are equal.
For antineutrinos, all curves show almost complete polarization reversal, with the
exception of small energies (of few MeV, not shown in Figure 3). 
Figures~\ref{fig3} and \ref{fig6} show the evolution of $J$ and
 $J_z$ for neutrinos and antineutrinos, in the single- and multi-angle cases.
The behavior of these vectors can be related to the gyroscopic pendulum motion. 
At the beginning, in the synchronized regime, all the
polarization vectors are aligned so that  $J=J_z$ and $\overline J=\overline J_z$: the
pendulum just spins in the upward position without falling. Around $\sim 70$ km
the pendulum falls for the first time and nutations appear. The nutation amplitude gradually decreases 
and bipolar oscillations eventually vanish for $r\sim 100$~km. At the same time, the spectral split builds up:
antineutrinos tend to completely reverse their polarization, while this happens only partially for
neutrinos. As said before, also for antineutrinos there is a partial swap of the spectra for $E\sim 4$~MeV.
From Figure~\ref{fig6}  it appears that 
bipolar oscillations of $\bf J$ and $\overline{\bf J}$ are largely smeared out in the multi-angle case.
The bipolar regime starts somewhat later with respect to the single-angle
case,  since neutrino-neutrino interaction angles can be larger than the (single-angle) average one,
leading to stronger self-interaction effects, 
that force the system in synchronized mode  slightly longer.
However, just as in the single-angle case, the spectral split builds up, $\overline{J}_z$ gets finally reversed, while
the difference $D_z=J_z-\overline{J}_z$ remains constant.
\begin{figure}[t]
\psfig{figure=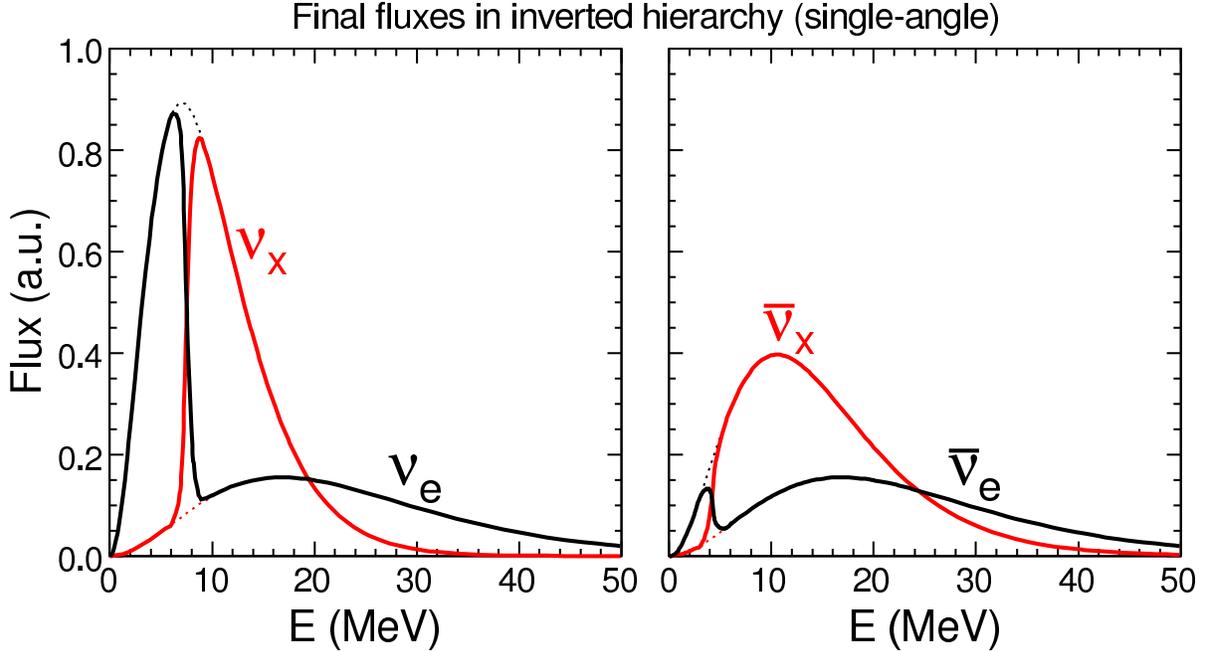,height=3.4in}
\caption{Single-angle simulation in inverted hierarchy: 
final fluxes (at $r=200$~km, in arbitrary units) 
for different neutrino species as a function of energy. Initial fluxes are shown
as dotted lines to guide the eye.
\label{fig5}}
\end{figure}
Figures~\ref{fig5} and \ref{fig8} show the final neutrino and antineutrino fluxes, in 
the single- and multi-angle simulations.
The neutrinos clearly show the spectral split effect and the 
corresponding sudden swap of $\nu_e$ and $\nu_x$ fluxes above $E_c\simeq 7$~MeV.
In the right panel of Figure~\ref{fig5}, the final antineutrino spectra are basically
completely swapped with respect to the initial ones, except at very low energies, where there 
appears an ``antineutrino'' spectral split.
This phenomenon can be related to the loss of $\overline J$ 
and of $|\overline{J}_z|$~\cite{ourCollective}.
Also in the multi-angle case of Figure~\ref{fig8} ,
the neutrino spectral swap at $E>E_c\simeq7$~MeV is rather
evident, although less sharp  with respect to
the single-angle case, while the
minor feature associated to the ``antineutrino spectral split''  is largely smeared out.

\section{Conclusions}

We have studied supernova neutrino oscillations in a model where the collective flavor transitions
(synchronization, bipolar oscillations, and spectral split) are
well separated from the MSW resonance.
We have performed numerical simulations in both single- and multi-angle
cases, using continuous energy spectra with significant $\nu$-$\overline \nu$
and $\nu_e$-$\nu_x$ asymmetry. The results
of the single-angle simulation can be analytically understood
to a large extent by means of a mechanical analogy with the spherical pendulum. 
The main observable effect is 
the swap of energy spectra, for inverted hierarchy,
above a critical energy dictated by lepton number conservation.
In the multi-angle simulation, the details of
self-interaction effects change (e.g., the starting point of bipolar oscillations and their amplitude),
but the spectral swap remains a robust, observable feature. In this sense,
averaging over neutrino trajectories does not alter the main effect of the self interactions. 
The swapping of neutrino and antineutrino spectra could have an impact on
$r$-process nucleosynthesis, on the energy transfer to the shock wave during the supernova
explosion and on the propagation of the neutrinos through the shock wave.
From the point of view of neutrino parameters, collective flavor oscillations in 
supernovae could be instrumental in identifying the inverse neutrino mass hierarchy, even for very small 
$\theta_{13}$.~\cite{DigheMiri}
 \begin{figure}[t]
\psfig{figure=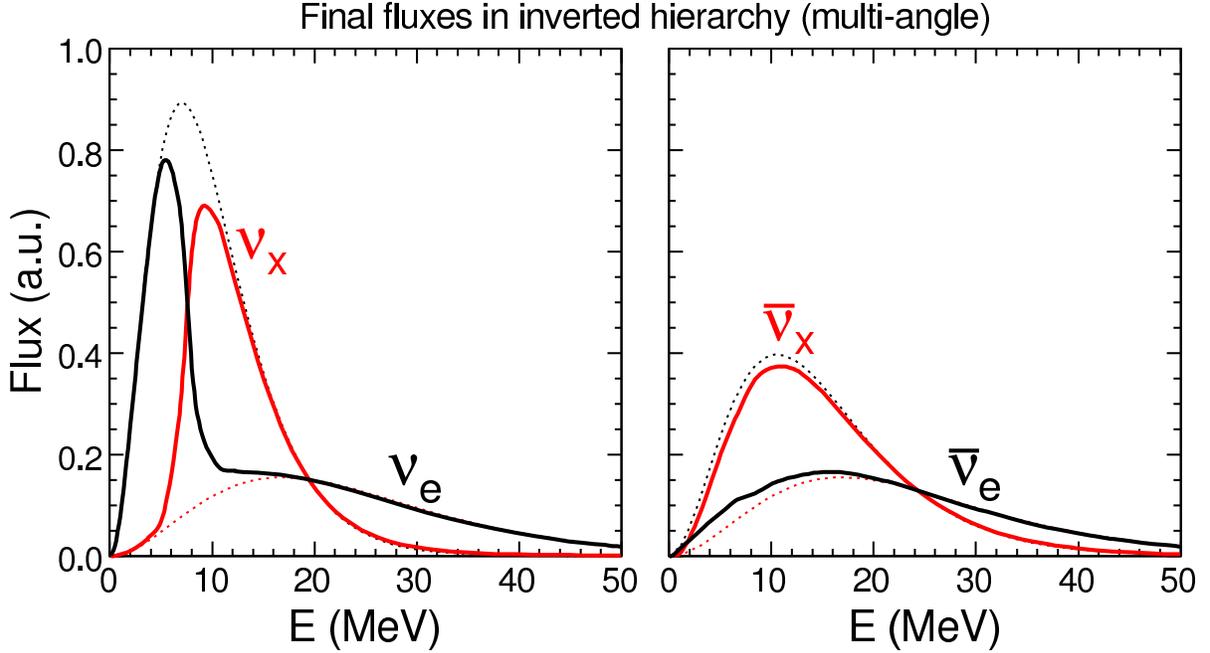,height=3.4in}
\caption{Multi-angle simulation in inverted hierarchy: final 
fluxes (at $r=200$~km, in arbitrary units) 
for different neutrino species as a function of energy. Initial fluxes are shown
as dotted lines to guide the eye.
\label{fig8}}
\end{figure}

\end{document}